# Semantically Enhanced Time Series Databases in IoT-Edge-Cloud Infrastructure

Shuai Zhang, Wenxi Zeng, I-Ling Yen, Farokh B. Bastani
Department of Computer Science
The University of Texas at Dallas
Richardson, Texas 75080
{shuai.zhang1, wenxi.zeng, ilyen, bastani}@utdallas.edu

**Abstract**. Many IoT systems are data intensive and are for the purpose of monitoring for fault detection and diagnosis of critical systems. A large volume of data steadily come out of a large number of sensors in the monitoring system. Thus, we need to consider how to store and manage these data. Existing time series databases (TSDBs) can be used for monitoring data storage, but they do not have good models for describing the data streams stored in the database. In this paper, we develop a semantic model for the specification of the monitoring data streams (time series data) in terms of which sensor generated the data stream, which metric of which entity the sensor is monitoring, what is the relation of the entity to other entities in the system, which measurement unit is used for the data stream, etc. We have also developed a tool suite, SE-TSDB, that can run on top of existing TSDBs to help establish semantic specifications for data streams and enable semantic-based data retrievals.

With our semantic model for monitoring data and our SE-TSDB tool suite, users can retrieve non-existing data streams that can be automatically derived from the semantics. Users can also retrieve data streams without knowing where they are. Semantic based retrieval is especially important in a large-scale integrated IoT-Edge-Cloud system, because of its sheer quantity of data, its huge number of computing and IoT devices that may store the data, and the dynamics in data migration and evolution. With better data semantics, data streams can be more effectively tracked and flexibly retrieved to help with timely data analysis and control decision making anywhere and anytime.

## 1 INTRODUCTION

Due to the rapid advances of IoT technologies in recent years, many systems have incorporated IoT to improve their system operations. IoT systems, such as smart homes, smart buildings, smart cities, smart planet, smart farms, smart agriculture, smart factory, smart manufacturing, smart industry, smart parking, smart transportation, smart grid, etc., have been steadily developed and deployed.

One major use of IoT systems is for monitoring and fault detection and diagnosis (MFDD). Sensors are attached to networking capable boards, such as Arduino launchpad, Raspberry Pi, etc., and placed in the systems to monitor system behaviors and pass out the monitoring data via wireless networks. Many industrial companies and government organizations are adopting IoT for MFDD of their critical systems. For example, manufacturers set up IoT systems to monitor the machines and operations in the production lines to help with early problem detection and mitigation to ensure reliable and safe system operations. IoT based monitoring can also help reduce the likelihood of producing defective products and potentially improve product quality. Critical national infrastructures, such as electricity grids, water and waste water systems, etc., also use IoT for system health monitoring to achieve the goal of early detection of faults in the target system in order to exercise timely mitigation plans.

The IoT based MFDD systems are generally data intensive. A large volume of data are collected from a large number of sensors and these data streams flow continuously. Thus, data storage and data analysis are major issues to be considered. Big data analytics have been widely used in recent years to analyze the data collected from IoT systems to achieve better fault detection and diagnosis [1] [2] for the target system. Due to the computation intensive nature of big data processing, the analyses generally need to be performed in the cloud. However, channeling the data to the cloud may have significant communication latency. Also, a lot of the fault detection and diagnosis tasks need to be performed in real time. In response to such conflicts between timeliness and resource constraints, the new computing infrastructure that integrates IoT, Edge, Fog, and the Cloud have been developed. Preliminary computations that are required for timely decisions can be done in the nearby Edge gateways and Fog LANs (local area networks). Integrated analysis can be left to the Cloud and the more sophisticated and/or more accurate analysis results can then be channeled back to the Edge and IoT devices to provide improved decision making or to support changing trends.

As mentioned above, data storage for time series data can be a problem due to their continuously cumulating nature. In recent years, many time series databases (TSDB) have been developed [3] [4] [5] [6] [7] to handle the storage of continuous data streams. However, due to the recency of TSDB research, there are still rooms for improvement in TSDB designs. The first issue is how to effectively down-sample the older data so that the storage space can be saved while the useful information in the historical data can be retained. Though all TSDBs provide down-sampling support, but getting effective down-sampling functions is still a research challenge.

The second problem with existing TSDBs is the lack of semantic support. Most of the existing TSDBs are designed with supporting a single application system in mind. They focus on data storage performance while lacking data semantics support. Thus, the data in the database can only be interpreted by those who design the database schema. This leads to critical shortcomings in effective data discovery and processing. For example, many modern cyber physical systems (CPSs) are highly complex. When an IoT based monitoring system is used for MFDD for a complex system, current TSDBs cannot capture the relations between the sensor data streams based on the relations between the entities they are monitoring, making it challenging to analyze the potential relations between the data streams. Also, there are many similar cyber-physical systems deployed in different organizations or places for the similar goals. Integrated data analysis, i.e., analyzing data streams from similar systems, can help improve the accuracy of the analyses. Transfer learning [8] can also help bootstrap the fault detection process of a new system that is similar to some existing ones. Without proper data semantics, it will be very difficult to cross reference similar data streams and achieve integrated analysis or transfer learning. The data storage problem becomes more severe when we consider the IoT-Edge-Cloud infrastructure. A data stream may be migrated and processed in the huge infrastructure and, hence, it is difficult to know the whereabouts of the data, making data retrieval more challenging.

The third problem with existing TSDBs is that they do not model the events. All TSDBs provide very good support in storing data streams that are collected at a relatively steady pace, i.e., data arrive at a certain interval. On the other hand, event data, such as system up and downs, etc., come to the system sporadically. Storing data streams and supporting efficient retrieval for event-based data streams is as critical as for data streams with relatively fixed collection intervals.

In this paper, we focus on enhancing the semantics of time series databases (the second problem discussed above) for data-intensive IoT systems. In an IoT-based monitoring system, we need to label the monitoring data streams in terms of several important attributes, including which sensor has generated the monitoring data stream, which metric of which target system entity is being monitored by the sensor, what is the relation of the target entity to the target system, what measurement unit is used for the data stream, etc. (Note: target system and target entity are the system and the specific entity in the system being monitored.) Many of these attributes are not modeled by existing TSDBs due to the complexity in specifying them. For example, in order to specify which entity in a target system is being monitored, we need to be able to specify the system architecture and construct the model for the specification of system entities. We also need to specify the sensors and their relations to the entities in the target system.

In order to specify the metric being monitored, we need to properly define the metric and its relation to other metrics. In other words, we need to define a metric ontology so that each metric referenced in a monitoring system can be properly understood.

We have developed the techniques to address the semantic specification problem discussed above. Specifically, we have developed a Measurement Description semantic model which defines the System Ontology, the Metric Ontology, the Measurement Unit ontology, and the relations between them. System Ontology defines the hierarchies of entities in the systems of an application domain. The entities in the system include the sensors and the entities being monitored. Through the System Ontology, we can define the system architecture that properly describe the target system and use it to specify which sensor is monitoring which entity in the system and the correlations between the entities in the target system (which infers the relations of the corresponding data streams). The Metric Ontology defines various metrics that may be used in an application domain and their relations. The Measurement Unit Ontology defines the units that can be used for various measurement metrics in the application domain and their relations. Accordingly, data streams can be specified by linking to the specific metrics, the specific entities, and the specific measurement units.

Based on the semantic model, we have also developed a tool suite, SE-TSDB, to manage the semantic information and to enable semantic based monitoring data retrieval. Via SE-TSDB, domain experts can specify the System Ontology, the measurement metric ontology, and the measurement unit ontology. System architects can use the System Ontology to further define the specific system architecture for the monitoring system and the target system. The Reasoning Engine in SE-TSDB can derive new data streams from the existing ones based on the associated metric definitions. Thus, users can issue a semantic query to retrieve a data stream that is not in the TSDB but derived by the Reasoning Engine. Analysis programs with mismatching outputs/inputs can interoperate by letting the Reasoning Engine derive the alignment rules. Users can also retrieve data streams based on similarity-based matchmaking. SE-TSDB allows domain experts to define similarity rules and matching mechanisms on top of the semantic model. Thus, monitoring data streams that are collected for similar entities in similar systems can be retrieved by a semantic-based query and the Reasoning Engine can discover the matching based on the similarity rules. SE-TSDB can run on top of existing TSDBs to enhance their semantic capabilities.

We use cloud as an example case study to illustrate how the semantic model should be constructed and how a semantic-based query can be processed to obtain the desired data.

The problem of lacking proper specifications for the data streams stored in the TSDBs becomes much more severe when we consider the IoT-Edge-Cloud infrastructure. A huge number of data streams may be generated at the edge and they may flow dynamically in the IoT-Edge-Cloud infrastructure to support various data processing and control tasks and to achieve various performance related objectives. How to label these data such that applications can easily discover the desired data streams in order to use them to make informed control decisions or to conduct integrated analyses. To cope with this problem, we discuss how our semantic model for data stream specifications can be used and the additional mechanisms needed to help achieve effective data discovery in the complex IoT-Edge-Cloud infrastructure.

The layout for the rest of the paper is as follows. In Section 2, we discuss the semantic model for data stream specification. The SE-TSDB tool suite is discussed in Section 3. Section 4 briefly discusses the issues of data stream retrieval in IoT-Edge-Cloud Infrastructure and outlines the potential solution. In Section 5, we use cloud as an example to illustrate how to construct the semantics and how to use the semantics for data stream derivation. Brief reviews of the literature in TSDBs and in semantic models for IoT and sensors are discussed in Section 6. Section 7 concludes the paper.

## 2 SEMANTIC MODEL FOR TIME SERIES DATA SPECIFICATION

The lack of semantics in existing TSDBs makes it difficult to support advanced matchmaking, data tracking, data-oriented reasoning, semantic based data retrieval, etc. We build upper ontologies to provide a better semantic model for time series data specifications for the entire lifecycle of the data. Our model mainly focuses on the specification of MFDD time series data. Thus, we first have to consider the basics for describing the monitoring data, including (1) which sensor is used for monitoring, (2) which entity in which system is being monitored and what is the relation of the entity to other entities in the system, (3) which metric is being observed for the entity, and (4) which measurement unit is used for the metric. To avoid having to construct the specifications above for each case, which will incur a lot of duplicate efforts, we can consider these specifications for an application domain. For (1) and (2), we can build a System Ontology to describe the potential entities in a system in an application domain and the common relations between these entities. A system architecture can be instantiated from the System Ontology for each specific system. Similarly, for (3) and (4), we can build the Metric Ontology and Measurement Unit Ontology for the application domain. When specific metrics are monitored and specific measurement units are used for a specific system, these ontologies can provide the domain-specific vocabularies to ease semantic specifications and to provide better understandability of the data. Figure 1 shows the upper ontology for domain-specific measurement data description.

In the figure, the black link is the "has" relation. The red link has two directions and represents two relations (to simplify the drawing), in which the right direction is the "has" relation and the left arrow is the "contains" relation. Different node colors simply show different categories of the nodes, and there is no differentiated ontological meaning. The green and the gray-green colored classes are from the specifications of the data stream.

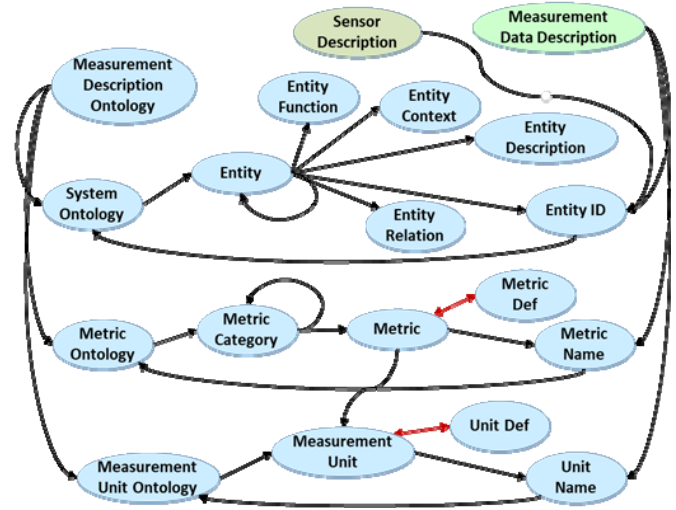

**Figure 1. Upper ontology for measurement description.**

As discussed earlier, the measurement description can be defined for an application domain and it includes the Metric Ontology, Measurement Unit Ontology, and System Ontology. These ontologies need to be defined by the domain experts. The System Ontology has entities in multiple layers. An entity may have some sub-entities. Some specifications can be given to each entity. A metric category organizes the related metrics together. Each metric has a name and a definition. A metric definition may contain other metrics (defined by these other metrics), which can be specified by the "contains" relation. Similar to a metric definition, a measurement unit may be defined by other units and the "contains" relation indicates that dependency. Note that the metric and unit definitions can be very useful in reasoning for matchmaking, data conversion between different application programs using different metrics and units, data stream derivations, etc. The System Ontology can help derive the property of a higher-level entity (category) from the properties of the lower-level entities.

The specifications about an entity include several categories of information as shown in Figure 1. The entity identity related information could include the entity name (maybe a path name in a hierarchical system structure to assure the uniqueness in the system), some identification number(s) (e.g., for a vehicle, there are vehicle identification number and license number), the category or type of the entity, entity manufacturer and manufacturing time, owner of the entity, etc. The entity relation specifies the connection and interaction relations of the entity with other entities in the system. The entity context could include location (can be static or dynamic over time), status over time, etc. The specifications related to the entity function can include the general functionality description, the specific functionality of the entity in the system, some general descriptions of the entity, etc. For some actuator entities, the precondition, effects, and behavioral descriptions can be given in its function description. The general description about the entity can include any other information that are not specifically categorized in the entity description ontology.

The Sensor Description (the gray-green node) is from the Provenance class of the data stream specification, specifying which sensor is used for monitoring. The Measurement Data Description (the green node) is from the data stream specification. It needs to specify a specific metric in the Metric Ontology and a specific target entity in the System Ontology that the sensor is monitoring and the specific measurement unit in the Measurement Unit Ontology the measurement uses.

Next we introduce the upper ontology for data streams (as shown in Figure 2), which defines the important elements for the specification of the time series (data stream).

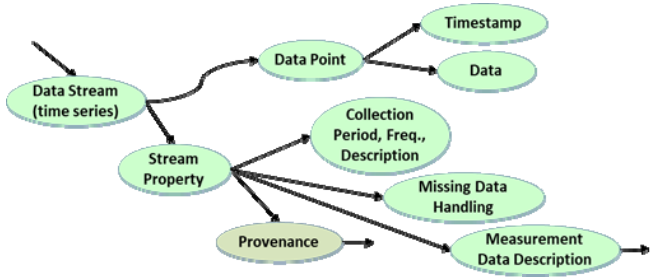

**Figure 2. Ontology for data stream semantics.**

A data stream includes a sequence of data points, each has the measurement data and a timestamp indicating when the data was collected. Other descriptions about the data can be defined at the data stream level to reduce the overhead. These descriptions can include the timing information such as data collection frequency and period, description about the collection procedure (such as the collection method, etc.), how to handle the missing data (such as using interpolation or just ignore it). The Measurement Data Description is an important class for data stream specifications, and its relation to the entities in the Measurement Data Description ontology have been given in Figure 1.

Another important descriptor in a data stream is the provenance information for the data stream, i.e., how is the data stream collected or derived [9]? The upper ontology defining Provenance is given in Figure 3.

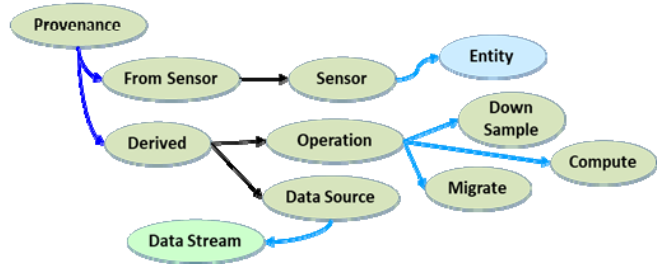

**Figure 3. Ontology for data source provenance.**

A data stream may be collected from a monitoring sensor or derived from other data streams. If a data stream is the raw sensor readings, then we associate the data stream to the sensor. Note that a sensor is an entity in the monitoring system. Thus, we use entity ontology to specify the sensor and its role in the system. If a data stream $s$ is derived, then we specify the input data sources, which are other data streams, and the operation performed to derive $s$ from the other data streams. The operation may be a migrate, down-sample, or compute operation. A compute operation can be specified in general by the computing function name and features and specifically by the url. Generally, provenance can be represented by a DAG (directed acyclic graph) in which the leaf nodes are the raw data streams and the internal nodes are the operations. With the recursive relation between provenance and the data stream, we can see how the DAG can be specified.

Finally, we define the upper ontology for TSDBs (Figure 4).

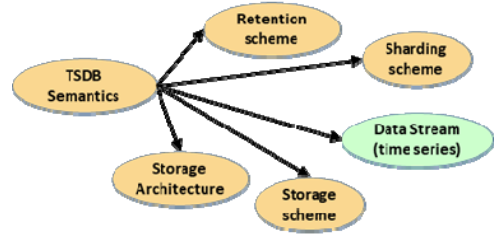

**Figure 4. The upper ontology for TSDB semantics.**

A TSDB basically includes a collection of correlated data streams. These data streams are generally from all the sensors for a target system or a subsystem. In addition to the data streams, the storage architecture (e.g., integrated IoT-Cloud-Edge), the specific retention and sharding policies, and the storage scheme can be specified. The storage scheme can include specific management mechanisms for performance, fault tolerance, and security.

## 3 SE-TSDB

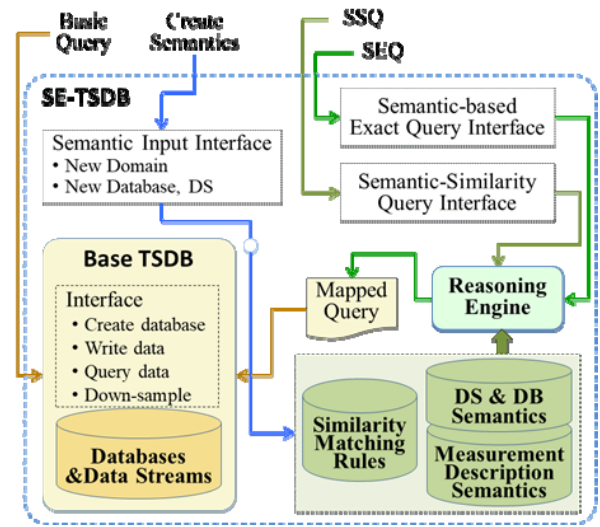

**Figure 5. SE-TSDB architecture.**

Existing TSDBs do not offer sufficient semantic description support for the time series data they store. We have defined a semantic model and corresponding ontologies to help build the enhanced semantics for time series data specifications. We have also developed a suite of tools to realize our semantic model for TSDBs, namely, the SE-TSDB (the semantically enhanced TSDB) tool suite. SE-TSDB is implemented as a thin layer on top of existing TSDBs. The architecture for the tool suite is shown in Figure 5.

The detailed descriptions for some components in SE-TSDB tool suite are discussed in the following subsections.

### 3.1 Semantic Information Creation

To enhance the semantics of the Base TSDB, the SE-TSDB tool suite needs to support the specification of different levels of semantics. The Semantic Input Interface (SII) is responsible for interacting with the users to obtain various semantic specifications:

(1) When a new application domain is created, the domain experts can specify the Measurement Description Ontologies for the domain, which can be the basis for specifying the semantics of the monitoring data streams. The Similarity Matching Rules define the similarity metrics for various components in the system. Some basic rules can be offered by SE-TSDB. Domain experts should define domain specific matchmaking rules for various similarity attributes and SII will store them in the Similarity Matching rule base.

(2) A TSDB can be created for a specific monitoring system for a certain target system (or subsystem). During database creation, the system architecture for the target system and the monitoring system can be specified, which instantiates the System Ontology and includes the entities of the target system and the monitoring system, their relations, and their descriptions. Some other properties of the TSDB, such as the storage and processing architecture for the data and the specific scheme regarding how to distribute the data over the storage architecture (the storage scheme) should be specified in the TSDB semantics.

Besides creating the TSDB semantics, the database should actually be created in the Base TSDB system. SII simply uses the database creation APIs offered by the Base TSDB to perform this action. At the same time, the TSDB semantics defined for the database will be associated to the base database by its name. After database creation, we specify the retention and sharding policies and down-sampling schemes offered in the Base TSDB.

(3) Each monitoring sensor will generate a continuous data stream (DS), which will be written to the data stream (DS) of the corresponding database. The semantics for each data stream should be specified before writing the data points. In some Base TSDBs, each write operation for a data chunk specifies the database and the data streams [5] (since there is no semantics, the overhead is negligible). In some other Base TSDBs, the database and the data streams have to be created in advance [7]. But in either case, the user can use SII APIs to create the database and the DS semantics. The association of the DS semantic data with the DS data points stored in the Base TSDB can be done by the database name and some DS attributes (depending on the data model of the Base TSDB). For most of the Base TSDBs, the database name, the specific metric name, and the specific tags are used to uniquely identify a data stream. Thus, we store these attributes in the DS semantics repository to achieve the proper association.

For RRD [3], an early TSDB that is used by several cloud monitoring systems as the monitoring data storage solution, the association method discussed above will not work. Thus, we have modified the read/write operations in RRDTool to incorporate the semantic enhancement capabilities [10].

### 3.2 Query Processing

The basic queries (**BQ**) are those without going through the SE-TSDB tool suite and are routed directly to the Base TSDB.

Sometimes, semantic based data retrieval is desired. Consider a few cases: (1) An application requires a specific input data stream which is not directly available in the TSDB. But there are some data streams in the TSDB that can be used to derive the desired data stream. For example, if the server status data (up/down status) is available in the TSDB but the user wants the availability data, then availability of the server can be derived from the up/down events and their timestamps using the Metric Ontology definitions. If a data processing program takes an input data stream in one measurement unit while the data streams in the TSDB are in another, then the conversion could be reasoned based on the Measurement Unit Ontology. (2) A user may not know the specific DS names and, hence, will use a semantic query to get the desired data. (3) When a new MFDD is deployed for a target system, it will have a cold start period. A data analyst may want to retrieve MFDD data of other similar target systems to get a quick start using transfer learning techniques [8]. In a more general case, the data analysts may retrieve MFDD data from multiple similar target systems and integrate them in fault analysis to improve the MFDD capability and accuracy.

In case (1), logical reasoning can be performed on formally defined semantics to infer the needed data retrievals and potential data aggregations. We call this type of queries the semantic-based exact queries (**SEQ**) and route them to SEQI (Semantic-based Exact Query Interface). Here "exact" is relative to "similarity based" and the query is processed by logical reasoning without fuzzy similarity matching. For case (2), we need to perform similarity-based reasoning to match the semantics specified in the query with the semantics of existing data streams based on some fuzzy similarity rules. Case (3) also requires similarity-based reasoning to retrieve data streams from target systems that are similar. We call this type of queries the semantic-similarity based queries (**SSQ**) and route it to SSQI (Semantic-Similarity based Query Interface) for further processing.

SEQI extracts the semantics defined in the query and forwards them to the Reasoning Engine, which in turn, takes the DS&DB Semantics and the Measurement Descriptions to perform logic-based reasoning and convert the input query and derive the new query, i.e., the Mapped Query. The Mapped Query is then sent to the Base TSDB to obtain the query results.

SSQI extracts the semantics and matching criteria defined in the query and forwards them to the Reasoning Engine. The Engine takes the DS&DB semantics and matches them with the semantics of the input query based on the rules in the Semantic Matching rule base to determine the databases and data streams that are relevant to the query. These "similar" items are further processed by the Reasoning Engine for logic-based reasoning. Finally, the Reasoning Engine produces the Mapped Query for TSDB data retrieval.

### 3.3 Similarity Rules

Similarity Matching rules need to be defined to enable the processing of the SSQ queries. Among the semantic attributes for a data stream (discussed in Section 2), the one that is likely to be used for semantic-similarity based retrieval are the Sensor Description and Measurement Data Description. The latter specifies which metric of which target entity is being monitored (measurement unit is not very useful in similarity based retrieval). The semantics of an entity does not standalone and can be augmented by the semantics of the target system (or subsystem) and the relations of the entity to the system. Similarly, the sensor semantics can also be augmented by the semantics of the target system and the role of the sensor entity in the target system. Thus, a semantic-similarity based query (SSQ) can consider specifying the above semantic attributes to discover the desired data streams. Let $S_q$ denote the semantic information specified in an SSQ $q$, which can be characterized by the vector of relevant semantic attributes for matchmaking as follows:

$S_q = (S_q^{Sys}, S_q^{Entity}, S_q^{Metric}, S_q^{Sensor})$.

Here, $S_q^{Sys}$ and $S_q^{Entity}$ are the semantics for the target entity and the semantics of the corresponding system, respectively, specified in the SSQ. $S_q^{Metric}$ and $S_q^{Sensor}$ are the semantics of the metric and the semantics of the corresponding sensor specified in the SSQ.

A data stream $ds$ in the TSDB also has these semantic attributes and can be represented as

$S_{ds} = (S_{db}^{Sys}, S_{ds}^{Entity}, S_{ds}^{Metric}, S_{ds}^{Sensor})$,

where $S_{ds}$ is the semantics for data stream $ds$. Note that the system definition is given for each DB, not each DS. Thus we have $S_{db}^{Sys}$ instead of $S_{ds}^{Sys}$ in the vector, and $db$ is the database that contains data stream $ds$.

When assessing similarity between $S_q$ and $S_{ds}$, the similarity for each individual semantic attribute should be assessed separately. Then they can be aggregated.

Domain experts need to specify the similarity metrics for the four similarity attributes, i.e., define similarity($S_q^X, S_{ds}^X$), where $X$ can be $Metric, Entity, Sensor, System$.

For similarity($S_q^{Metric}, S_{ds}^{Metric}$), we can use the keyword-based similarity assessment methods with WordNet and other ontologies. Keywords can be extracted from metric name and metric descriptions. Formal verification methods should be used to match the metric definitions. Note that a metric $m$ may be derived from other metrics and, hence, during similarity matching for $m$, the metrics used in $m$'s definition in the Metric Ontology, say $m_1, m_2, \ldots$, should also be considered, i.e., if $S_q^{Metric}$ includes $m$, then it has to be expanded to $m \vee m_1 \vee m_2 \vee \ldots$ to facilitate additional reasoning.

For similarity($S_q^{Entity}, S_{ds}^{Entity}$), entity attributes such as entity function, entity description, etc. need to be considered. Keywords can be extracted from these descriptions and similarity metrics can be defined using keyword based methods.

For similarity($S_q^{Sensor}, S_{ds}^{Sensor}$), since sensor is an entity in the system, the same approach as similarity($S_q^{Entity}, S_{ds}^{Entity}$) can be used.

Assessing similarity($S_q^{Sys}, S_{db}^{Sys}$) may be more complex than assessing similarity of other semantic attributes. The system specification in query $q$ can be the system architecture or simply the keywords. In the latter case, keywords can be extracted from the system descriptions in $S_{ds}^{Sys}$ to match with those given in $S_q^{Sys}$. For some retrievals that require more precision, such as for integrated MFDD, the complete system architecture can be given in $S_q^{Sys}$. In this case, $S_q^{Sys}$ and $S_{ds}^{Sys}$ can be viewed as graphs with entities as the vertices and relations between entities as the edges. We need to align the two graphs and identify matching vertices and edges. Graph edit distance can be used to evaluate the graph structural similarity [11] [12]. Similarity for the entity pairs and the edge pairs in two systems can be computed and integrated with the structural similarity of the system graphs. Domain experts can specify the graph based similarity metrics. Keyword based similarity metrics can be used for entity pairs and edge pairs.

Some attributes of the semantic data for $q$ or for $ds$ could be missing. When there are missing semantics, the matching may be less precise. For example, if $S_q^{Entity}$ is specified but $S_q^{Sys}$ is missing, then all the measurement data streams for the specific entity (and similar entities) will be selected even if they may be from very different systems.

### 3.4 The Reasoning Engine

The Reasoning Engine is responsible for semantic-based matchmaking. It includes four major reasoning components, namely, entity-based reasoning, metric-based reasoning, measurement-unit based reasoning, and similarity-based reasoning.

**Measurement-unit based reasoning** focuses on data conversion due to inconsistency in measurement units. The Reasoning Engine first checks the consistency between the required measurement units in the retrieving query and the actual units in the to be retrieved data streams. If conversion is needed, it retrieves the measurement unit definitions in the Measurement Unit ontology and derives the rule to convert the data to match with the unit given in the query. A Mapped Query will then be generated, which retrieves the designated data stream and performs data conversion on it.

**Metric-based reasoning** derives matchmakings based on the definitions of the metrics. In case a desired data stream does not exist in the Base TSDB, the Reasoning Engine retrieves the metric definitions in the Metric Ontology and determines whether the desired data stream can be derived from other data streams in the Base TSDBs. If so, the Reasoning Engine reasons the mapping rules and convert them into the Mapped Query which will retrieve the corresponding data streams in the Base TSDB and derive the desired output data stream. For example, the data for the availability metric of a device may not be directly available in the monitoring database. Based on the availability metric definition in the Metric Ontology, the Reasoning Engine generates the Mapped Query to retrieve the relevant device status (up/down times) data stream to derive the availability data stream.

**Composition reasoning** is used for the case when multiple data streams for different entities need to be aggregated to obtain the desired data stream. For example, consider a cloud system. When the requester wants to query the "load" of a cluster entity, the Reasoning Engine will first, based on the System Ontology, determine that a cluster includes a set of hosts and then, based on the system architecture, determine the specific hosts in the specific cluster. Next, it retrieves the specific cluster load metric definition from the Metric Ontology to generate the derivation rule and convert it into the Mapped Query. The query includes which host load data streams to be retrieved and how to aggregate them into the load of the cluster.

**Similarity-based reasoning** is only used for SSQs. If the Reasoning Engine receives the query from SSQI, then it will start with similarity-based reasoning using the similarity metrics defined in Similarity Matching rule base. The Reasoning Engine go through each database $db$ to compute the similarity of $S_{db}^{Sys}$ and $S_q^{Sys}$. If there is a good match, then the Engine proceeds to each DS in $db$ and evaluate the similarity of the other semantic attributes.

Since there may be a large number of DBs and DSs, it is not feasible to explore all the DBs and their DSs during reasoning. Thus, a preliminary filtering will be performed to eliminate some candidates. DSs in the TSDB are clustered into a tree based on the keywords from the DSs and their container DBs. The filtering process starts from the root to eliminate the unlikely branches. The criteria for pruning a branch at a higher level of the tree will be stricter, i.e., requiring a higher degree of mismatches. After pruning, the remaining set of DSs will be used for similarity reasoning. After similarity-based reasoning, the DSs passed the similarity evaluation will go through the other three reasoning processes.

We have developed the Reasoning Engine [10] [13] based on Jess [14], a backward chaining reasoner in Java. The definitions in the Measurement Description ontologies, the semantic information for the databases and data streams, and the similarity matching metric definitions are converted automatically into rules and facts in Jess rule language. Jess engine then performs reasoning accordingly to derive the matchings. Our engine then translates them into Mapping Query for the Base TSDB.

## 4  TSDB IN THE IOT-EDGE-CLOUD INFRASTRUCTURE

IoT devices are being deployed at an increasing rate and it is estimated that there are tens of billions of physical things that are connected to the Internet, and the number is still growing rapidly. Many of these IoT devices are sensors, which generate large volumes of data. Generally, IoT devices have limited computing power and storage space and, thus, they are not suitable to store and process the data continuously generated by themselves. Cloud provides enormous computing and storage capacities which can be used for IoT data processing and storage. However, many IoT systems collect data for real-time decision making, and the communication latency for transferring data to the centralized cloud data centers can be prohibitive. To resolve these tradeoffs, edge and fog computing solutions have been proposed. Edge nodes generally serve as the gateway for the IoT devices and can perform some preliminary computations. Fog is a mini-scale cloud that sits between the Cloud and the IoT edge and offers computing resources, data storage, and networking in a local area network (LAN). An IoT-Edge-Fog-Cloud infrastructure is illustrated in Figure 6. In subsequent discussions, we merge Edge-Fog to one Edge entity for simplicity.

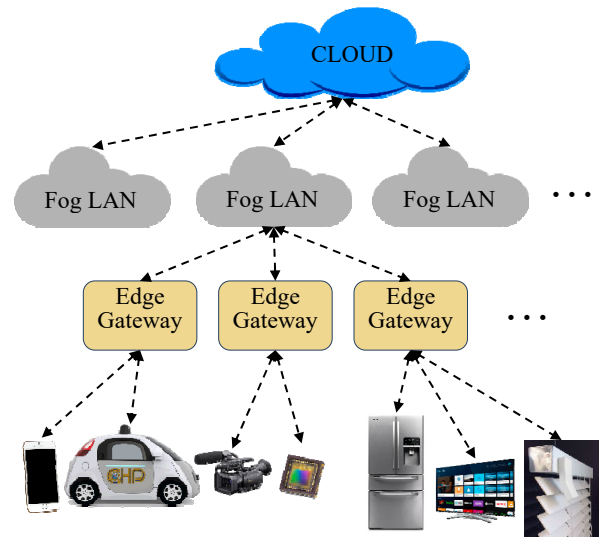

**Figure 6. IoT, edge, fog, and cloud computing infrastructure.**

The IoT-Edge-Cloud infrastructure constitutes a huge-scale distributed platform for IoT data processing and storage. To balance the tradeoff between timeliness and storage capacities, we can let IoT devices host transient data that are needed for quick decision making and let edge and cloud host persistent data. Data streams from IoT devices can flow to edge for preliminary processing. The original data or processed data can further flow to the cloud for integrated processing and potential storage. TSDBs can be used for data storage in the infrastructure. Each edge node can run the single node TSDB instance and manage the data as a peer. The cloud can run a distributed TSDB instance as a super peer.

A major problem with the huge distributed storage in the infrastructure is how to manage the data. For example, an application may wish to retrieve the current traffic volume of some streets to make a driving plan. A user may find some analysis results of faulty behaviors of various computing systems and wish to find the original data sources for the analysis to gain further insights. A system designer may wish to obtain the failure information of a similar system to help evaluate the fault tolerance features in her system design. How should these retrievals be processed in the IoT-Edge-Cloud infrastructure. As can be seen, maintaining the semantic information about each data stream can greatly help correlate, locate, and retrieve the data streams in the infrastructure.

We consider data stream retrieval in three modes:

(1) Localized retrieval. If we know where the desired DSs are hosted or we only plan to retrieve DSs from a specific TSDB, then the three types of retrieval queries, BQ, SEQ, and SSQ, discussed in Section 3.2 can be used on a specific TSDB system to obtain the desired data streams.

(2) Infrastructure-wide semantic-based discovery. If we do not know where the desired DSs are hosted or we would like to perform a general search to discover all matching DSs in the whole or a part of the infrastructure, then we can issue an infrastructure-wide query. The query could be an SEQ or an SSQ and the desired semantics should be specified in the query for matchmaking. A region parameter can also be included in the query specification to confine the discovery in a specific region or in the entire infrastructure.

(3) Provenance based data discovery. Sometimes, users may want to track down the source of data for a derived DS or vice versa. Or, a time period of a data stream may be migrated from its source to other locations for processing. The provenance information maintained for a data stream can help identify the desired data streams in the infrastructure. Provenance data are stored in decentralized Provenance Data Repositories (PDRs). To allow a relatively stable provisioning of the provenance data, we assume that the PDRs are relatively stable, i.e., the provenance data are unlikely to be migrated from one PDR to another. Thus, we can store the repository reference address in the provenance data to link to the corresponding repository.

Infrastructure-wide discovery may include two types of searches, the discovery of a specific data stream (find out where it is hosted) with some uniquely identifiable semantic information and the general semantic based discovery to identify data streams satisfying the provided constraints.

Our DS semantic model has been designed to include information that can uniquely identify a data stream. Generally, a DS is far more dynamic than the sensor the DS is generated from and the target entity the sensor is monitoring. Thus, our semantic model incorporates information that can be used to uniquely identify entities (the entity semantics discussed in Section 2, include the location of the entity). Accordingly, a raw data stream can be uniquely identified by its sensor and target entity. If all the raw data streams can be uniquely identified, then all the data streams derived from them can be uniquely identified. However, missing semantic information should always be allowed and it can impact the capability of uniquely identifying the corresponding entities. In this case, we will simply use general semantic based matchmaking to identify the data streams satisfying the semantics specified in the queries.

To facilitate infrastructure-wide discovery, whether it is for discovering a specific DS or it is a general semantic based retrieval, we need to provide some structured information to guide query routing. We consider using peer-to-peer information structure with super peers maintaining the routing information for subsidiary peers [15]. In semantic based retrieval, how to design the summary information structure for routing can be challenging due to the high dimensionality of the semantic information attributes in our model. We plan to use the ontology coding technique discussed in [16] to greatly reduce the routing information to be maintained at the super peers. The specific routing table and routing algorithm designs for our DS/DB semantic model will be left for future research.

## 5 CASE STUDY: THE SEMANTIC MODEL FOR THE CLOUD

Cloud computing has rapidly evolved and become a prevailing paradigm, but behind the convenient cloud provisioning are the increasingly complex cloud infrastructures and services that require great management efforts. To assure the healthiness of the cloud and the quality of the cloud services, cloud monitoring, fault detection and diagnosis have been a pressing issue in many cloud systems. A lot of cloud monitoring systems have been developed in industry and academia, but most of them do not consider how to effectively manage their monitoring data. The monitoring data are generally stored in proprietary repositories or existing databases (such as RRD [3]). Frequently, cloud monitoring data can only be interpreted by those who store them. Here, we use cloud monitoring as an example case to illustrate how to construct the semantics for the monitoring data and how our semantic model can enhance the usability of the data.

### 5.1 System Ontology for the Cloud

We reference the cloud concept categorizations in [17] and [18] and the general cloud infrastructure to build the System Ontology for the Cloud and it is depicted in Figure 7.

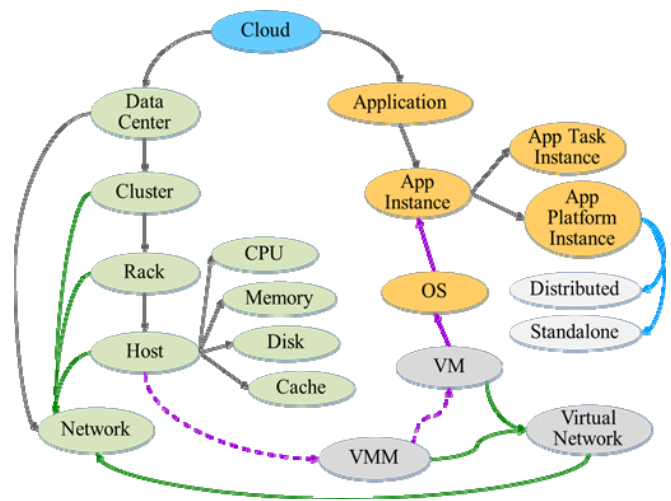

**Figure 7. System Ontology for the Cloud.**

A cloud has a hardware infrastructure as well as applications that are cloud services and/or for apps of individual customers. On the left side of the ontology (left of VMM), the cloud hardware infrastructure in a data center is defined, from host organized in racks to clusters in a data center. These hardware entities are linked together by the network. On the right side of the ontology, the software layers are given, including the VMM hosting VMs, the OS in the VM, and the application instances running on top of the OS. The VMs and VMMs may communicate via virtual as well as physical networks. Each cloud application may have one or more App Instances, which may be distributed over multiple hosts in the cloud or running on a single host. An App Instance can run directly on the OS or on top of some application platform. For a distributed application, the app platform needs to manage application task distribution, communication, synchronization, etc. A system ontology instance (the system architecture) should be created for each real cloud system based on the ontology. Each entity in the ontology can be described following the entity ontology defined in Figure 1.

Cloud monitoring sensors can be deployed to collect monitoring data for each entity in the cloud system. Some of them are hardware sensors and others are software sensors. There are standardized monitoring sensors for cloud hardware and host supporting systems (VMM, VM, OS), but the sensors for the application instances are application dependent and should be provided by the applications.

## 5.2 QoS Metric Ontology for the Cloud

QoS ontology has been explored in the literature. It defines the QoS metrics and their relations. Our QoS ontology is based on the works of [19], [20], and [21]. We integrate these works to build our basic QoS ontology. However, besides the QoS metrics themselves, the metric ontology should also include information specifying each metric. Our metric ontology model is depicted in Figure 8.

The "hasMetric" relation is the main relation in the QoS metric model. It links various QoS metrics together in a class hierarchy. With only the hasMetric relations, our ontology has the same model as other QoS ontologies. In addition to the hasMetric relation, our QoS metric model adds three additional subclasses to describe the metric. The subclass MetricDescription gives the natural language description of what the metric is. Since different names are used in different QoS ontologies for the same concept, we define a ConceptPool class for each QoS metric to provide various literal definitions for the same concept. For example, "CPUTime" has a concept pool of "CPUCredit" [22]. The purpose of the concept pool is to facilitates integration and interoperation of multiple ontology definitions, in lieu of the unavailability of technical terminologies in common linguistic ontologies such as WordNet. The Quantitative Definition class provides a mathematical definition of the QoS metric based on other metrics and/or the timestamps of those metrics Thus, a QoS metric may "contain" other QoS metrics in its quantitative definitions as specified in the System Ontology in Figure 1.

A partial view of our QoS metric ontology is given in Figure 8. In Figure 8(a), we can see that QoSMetricConcept includes a set of high level metric classes. Each high level class can be expanded into submetrics along the hasMetric relation. Figure 8(b) shows the expanded view, i.e., the submetrics for the Performance metric class. We use the annotation feature provided in Protege to specify the other three descriptive classes (linked to the metric by the hasDescription relation) for each metric. The sample description class for the "Availability" metric is given in Figure 8(c). It includes description and quantitative definition. Since availability metric is very standard, there is no conceptPool annotation for it.

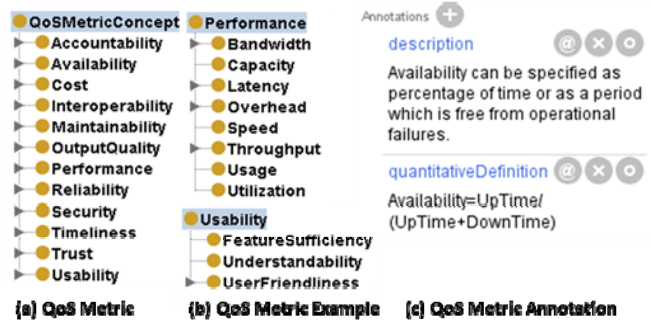

Figure 8. Partial view of the QoS metric ontology for the Cloud.

## 5.3 Measurement Unit Ontology for the Cloud

The measurement unit ontology is a class hierarchy of units of measurements (a partial view is shown in Figure 9(a)). At the highest level, the measurement units are classified into basic units and aggregate units. The classes of basic units are enumerated in Figure 9(b). The aggregate unit class includes units that are defined by other aggregate and basic units through some mathematical operations. It is further divided into three classes, ratio unit, volume unit, and complex aggregate-unit classes (as shown in Figure 9(a)). Speed, frequency, throughput, etc., are examples of ratio unit. An aggregate value obtained from multiple units through multiplication operation has a volume unit. An example volume unit is VM demand volume (combination of CPU, memory, disk requested).

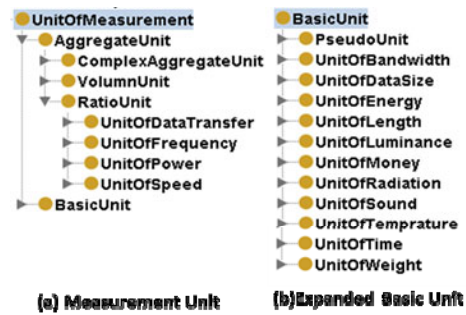

Figure 9. Partial view of the Measurement Unit Ontology for the Cloud.

## 5.4 Example for Semantic-based Reasoning

Consider "cluster availability" as an example. Cluster availability is derived from the availability of all host machines in the cluster. Hence, it is necessary to understand the relation between the concepts of "cluster" and "host", which is available in the system ontology instance for the specific cloud. Also, availability may not be measured directly and the availability data stream should be derived from the timestamps of the entity failures and entity recoveries stored in the status data stream of the entity.

In the query for cluster availability, a specific cluster and the availability metric are specified. Our Reasoning Engine first performs composition reasoning to map a cluster to the corresponding hosts based on the Cloud System Ontology instance and generates the query that will retrieve the availability data streams of individual hosts and aggregates them according to the cluster availability metric definition. The individual host availability query will then be processed by the Reasoning Engine based on the availability definition in the Cloud QoS Metric Ontology for availability. The host availability query is mapped into a query which retrieves the up/down status data stream of the given host and computes the availability of the host accordingly. All the mappings results will be integrated by the Reasoning Engine into the final Mapped Query for the Base TSDB (currently it is the RRD).

## 6 LITERATURE SURVEY

Owing to the rapid development of IoT, huge amount of data flows from sensors to databases on a daily basis. This trend yields increasing demands on the scalability and usability of data storage solutions for handling time series data. Time series databases (TSDBs) are, thus, rapidly emerging and advancing. TSDBs have several different aspects compared to conventional relational or no-SQL databases. Due to the continuous flow of data streams, it is necessary for TSDBs to incorporate mechanisms to process staled data. Due to the additional dimension of time, TSDB needs to consider optimal storage to efficiently handle new type of queries that involves time criteria. Time based interpolation and alignment are also additional considerations in TSDBs due to its time dimension.

RRDtool [3], as the pioneer of TSDB, is a file-based database. It models data collected from different sensors as different data sources (DSs) and handles them independently. Also, it interpolates data to fit a predefined, fixed time interval. Graphite [4] adopts the data model of RRD but modifies the design of exact time interpolation with the use of timestamps for recording the actual data collection time. Though RRDtool and Graphite present an efficient solution to handle time series data, they have scalability problem due to their single-node based solution. OpenTSDB [5], like many other TSDBs, is a table-based (wide-column) database. It considers the situation that some correlated data streams may come to the system together and may be accessed together and, hence, allow users to define these data streams together in one table. Consequently, the write and read performance can be improved. However, table-based TSDBs have the problem of having to store redundant tags to differentiate the data streams. To compensate this space overhead, OpenTSDB introduces the Rowkey design which is essentially a compromise from table storage to DS based storage (like RRD), but only for a time segment. Thus, it achieves performance gain as well as space saving benefits. Although OpenTSDB offers high scalability, it is built on top of an existing distributed non-SQL database, which introduces an extra layer and incurs extra overhead. Thus, newer TSDBs build their cluster solutions from scratch. RiakTS [6] and InfluxDB [7] are such databases. Though all of them have the wide-column table based data model, RiakTS allows the user to define the partition key to guide the application-specific sharding and potentially achieve better access performance. InfluxDB, on the other hand, focuses on indexing all the table attributes, such as measurement names, tags, and time, to facilitate efficient random-access query processing.

As the data continuously flow in to the storage system, it is impossible to keep expanding the storage infinitely. Thus, freeing up space for new data but at the same time allowing the retrieval of historical data is a crucial feature of TSDB. Different TSDBs provide various retention policies. RRD allows users to specify consolidation policies to consolidate old data into coarser grained archives. Also, it provides multiple level of archives for further consolidation. Similarly, OpenTSDB designs the downsampler to consolidate data. InfluxDB believes that to consolidate a data stream is just a special case of deriving a new data stream from an existing one. Thus, instead, it allows users to define continuous queries that are executed periodically to create new data streams from the existing ones. The original data are discarded after the retention duration.

Data sharding is needed in all distributed storage systems. Unlike no-SQL databases in which data are ordered by their keys, sharding for time series data has to consider the key and the time spaces. RRD and Graphite do not consider sharding. Sharding in OpenTSDB is handled by the underlaying distributed file system (together with their row key design). RiakTS allows the user to define the "partition key" to control how to shard the data. However, naïve users may give improper partition key design which can greatly hinder the system performance. InfluxDB allows the user to specify a time range for sharding. Multiple data streams in one table within the user specified time quantum will be stored in one shard.

Though most TSDBs handle sharding and down-sampling issues, their considerations in data semantics are limited. In RRD and Graphite, the data source (DS) name is the only semantic information to describe each data source. The table-based TSDBs [5] [6] [7] support a better semantic description for the data streams than RRD by using multiple attributes (metric name and tags). But the flat attribute schema does not offer relational information among the attributes. Also, tags are used for differentiating data streams and cannot be used to add additional semantic information. Lacking a good semantic model in existing TSDBs implies their inflexibility in data retrieval. The retrieval queries have to provide the specific database and data stream names. It is also difficult to track the data streams after their migration, consolidation, or evolution. Our SE-TSDB is designed targeting these shortcomings to enhance the semantics of existing TSDBs.

Though existing TSDBs do not support semantic definitions, there are research works that define ontologies relevant to monitoring data streams. The sematic sensor network (SSN) [23] defines a specification model for sensors and later extended to actuators. Fiesta-IoT [24] and IoT-Lite [25] generalize SSN from sensor specific to the generalized "Device" concept. Fiesta-IoT merges sensor and actuator ontologies in SSN into one and further clarified their relations to the System class. These ontologies can be used for the specification of the "entities" in our System Ontology of the Measurement Description Ontology. In SSN, etc., the concept of "System" focuses on platforms and deployment, not the correlations of entities that constitute the system. Thus, cannot be used to derive the properties of a higher level entity from the properties of the lower level entities. The Observation class in SSN is defined for the sensor and Fiesta-IoT modified it so that the Observation becomes the central entity that uses a sensor for data readings. Thus, it is similar to our Data Stream class, except that our data stream class is data-centric and, hence, include more specific information about the "data" (such as Provenance, etc.) and the "storage" (such as the retention scheme, etc.) concepts. The observation in Fiesta-IoT has "QuantityKind", which is similar to our "Metric" concept. But our Metric is described by a Metric Ontology, which can facilitate derivation of one metric from other metrics and it is not supported by QuantityKind in Fiesta-IoT.

There are some concrete ontologies that can be very helpful in data stream specification. Specific metric ontologies, such as QoSOnt [26] and DAML-QoS [20] which define QoS metrics for service-based systems and the Trustworthiness ontology in [21] which defines the reliability, security etc. attributes, can be incorporated as a part of the metric ontology in the computing domain. The measurement units ontologies, such as M3-IoT [27] which defines a relatively complete taxonomy of metrics that current IoT sensors measure and UCUM [28] (unified code for units of measure) which defines the units of measures commonly used in science, engineering, and business fields, can be used to help build the metric and measurement unit ontologies in the corresponding domains.

In general, the semantic solutions discussed above are not specific for data stream specifications and, hence, can only offer partial solutions under the overall ontology discussed in this paper. Also, our SE-TSDB tool suite can support the semantic based derivations instead of simply providing the specification solutions.

# 7 CONCLUSION

In this paper, we consider the data storage issues in IoT-based monitoring systems, especially those for fault detection and diagnosis of critical target systems. Problems in existing TSDBs in terms of their lack of semantic descriptions have been identified. A Measurement Data Description semantic model has been developed to support the proper specifications of monitoring data streams and to enable semantic-based data retrieval. We have also developed the SE-TSDB tool suite, which can run on top of existing TSDBs to add the semantic power to them. We use the cloud system as a case study to illustrate how to construct the detailed semantic models and how a reasoner can use the semantic information in our model to reason for semantic based queries.

Our future research in SE-TSDB include several directions. First, we plan to define a set of Semantic Matching Rules to enable more powerful reasoning of semantic-based queries. We have defined the similarity-based matchmaking principles. We will encode them into rules and evaluate their effectiveness in realistic scenarios.

We also plan to expand our approach to achieve semantic-based data stream discovery in the IoT-Edge-Cloud infrastructure. Specifically, we will consider the peer-to-peer system structure with super-peers and explore various summarization techniques for the super-peers to build routing tables. The goal is to achieve efficient query routing in order to quickly discover the desired monitoring data.

Moreover, as discussed in Section 4, we sometimes need to discover data streams using the provenance information associated to the data streams. We plan to extend our work in integrated data provenance and information flow control [29] to establish the provenance information for monitoring data streams. We will validate the provenance data model defined in this paper and examine its effectiveness in tracking data sources in the IoT-Edge-Cloud infrastructure. We will also develop tools to support effective provenance-based data stream discovery.

# 8 ACKNOWLEDGEMENT

This research is supported in part by the National Science Foundation (NSF) Industry/University Collaborative Research Center (I/UCRC) Award No. IIP–1361795.